\documentclass[11pt]{article}
\usepackage[letterpaper,margin=1in]{geometry}
\usepackage[T1]{fontenc}
\usepackage[utf8]{inputenc}
\usepackage{lmodern}
\usepackage{amsmath}
\usepackage{graphicx}
\usepackage[colorlinks=true,linkcolor=black,citecolor=black,urlcolor=blue]{hyperref}

\title{Strength in Numbers: How Widely Shared Preferences Outweigh Stronger Ones in Football's Player Awards}
\author{Guanyu Hu\thanks{Michigan State University. Corresponding author: huguanyu@msu.edu} \and Xiao Wang\thanks{Purdue University}}
\date{November 2026}

\begin{document}
\maketitle

\begin{abstract}
\noindent When an expert jury picks a winner, which favoritism should organizers fear most: the strong loyalty of the few jurors who share a candidate's nationality, or the weaker fondness of the many who share the candidate's region? We studied 16,091 published ballots from world football's player of the year awards. Jurors favored compatriots, with odds multiplied by 3.7 to 13.1 per ranking step, and, more weakly, players from their own continental confederation. Yet when we replayed 40 contests with the estimated bonuses switched off, every changed leader came from the regional bonus rather than the compatriot one. In simulations, balancing the leading candidates' regional blocs cut the probability that favoritism decides a tied race from 0.38 to 0.12, while barring compatriot votes did almost nothing. Ballots cast by foreign national-team coaches show that national favoritism follows the coach's job rather than the coach's passport, while regional affinity does not move with the job at all. The replays are model based, and probabilities for individual contests move with the specification, but the central ordering does not: a strong preference shared by few jurors matters less than a weaker preference shared by many.
\end{abstract}


\section*{Media Summary}

Every year, the coaches, captains and journalists who vote for football's player of the year awards publish their ballots, and every year fans find hometown loyalty in them. Our study of 16,091 ballots asks a sharper question: which favoritism actually changes who wins? The answer is not the obvious one. Compatriot loyalty is the strongest bias on paper, but so few voters share a nationality with any candidate that it never tips the outcome. The quieter preference for players from one's own continent is shared by whole blocs of voters, and in close races those small pushes add up. Replaying 40 award votes with the biases switched off, every changed winner traces back to regional blocs. Simulations of alternative juries point to a practical fix: balance the leading candidates' regional blocs on the jury, rather than simply barring compatriot votes.

\section{A scandal of one judge, a bias of many}

At the 2002 Winter Olympics in Salt Lake City, a single French judge's score in the pairs figure skating final set off a scandal that forced the International Skating Union to abandon its century old scoring system. One judge, one mark, one overturned result. Stories like this teach us to watch for the juror with the strongest personal tie to a candidate, and sports journalism has not forgotten the lesson: every year, reporters comb the published ballots of FIFA's player awards for signs of hometown loyalty, and they rarely come away empty handed.

The suspicion is well founded. That evaluators favor candidates from their own group is one of the best documented facts about judgment: Olympic judges score compatriots higher than other judges do \cite{zitzewitz2006,emerson2009,sandberg2018,krumer2022}, reviewers of grants and manuscripts favor applicants from their own country \cite{fuchsschundeln2026,zumeldumlao2025}, and referees and juries favor members of their own group \cite{garicano2005,price2010,anwar2012}. The standard safeguard follows directly. Bar the conflicted juror. If your nationality matches a candidate's, you may not judge that candidate.

But the Salt Lake City logic, and the safeguard built on it, assumes that the strongest tie is the dangerous one. Whether a candidate actually wins depends on something else as well: how many jurors share a tie with that candidate. Compatriot favoritism is strong, but it is exercised one juror at a time, because a candidate has few compatriots on any jury. Regional affinity is weaker per juror, yet it is shared by a whole bloc of jurors at once, so its small pushes all point in the same direction and can add up. Figure~1 shows the arithmetic of this distinction. In FIFA's awards, a juror and a nominee share a nationality in only 0.5\% of pairings, but a continental confederation in 25\%. A rule that removes the first tie leaves the second one, the one that can add up, completely untouched. The theme of this article is that a strong preference shared by a few voters can matter less than a weaker preference shared by many, and that evaluation systems should be designed with both the strength and the reach of preferences in view.

\begin{figure}[t]
\centering
\includegraphics[width=\textwidth]{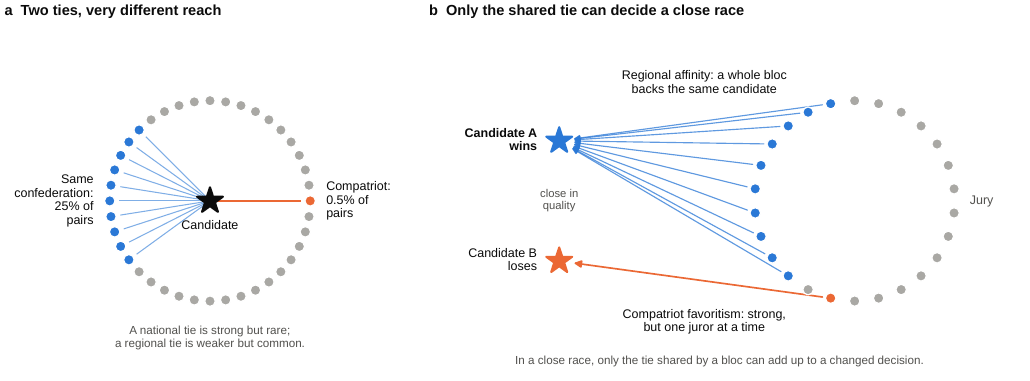}
\caption{Two ties, very different reach. (a) A candidate's ties to a jury are mostly regional, not national: in FIFA's awards a juror and a nominee share a nationality in 0.5\% of pairings and a confederation in 25\%. (b) Compatriot favoritism is exercised one juror at a time and rarely changes a collective ranking; regional affinity is exercised by a whole bloc at once and can tip a close race.}
\end{figure}

Which kind of favoritism actually changes decisions? World football's player awards let us answer the question directly, because their ballots are public. The closest prior work studied FIFA's ballots for the men's award in 2011 to 2014, documented compatriot favoritism among coaches, captains and journalists, and concluded that bias would change results only when the leading candidates were close \cite{coupe2018,coupe2016wp}; later work linked the preferences to cultural distance and jury composition \cite{johnson2022,kopkin2022}. In recent work \cite{huwang2026bdo} we went deep on a single award, letting journalists in the 2019 and 2024 Ballon d'Or vote in different evaluation styles while sharing one performance yardstick. Both races turned out to be nearly even contests between a European and a South American nominee, and in both the part of the bias with leverage on the outcome was the regional bonus, not the compatriot one. Here we widen the lens from two dramatic contests to sixteen years: both awards, both sexes, two jury designs, and every decision reconstructed from every ballot cast.

\section{Sixteen years of public ballots}

FIFA publishes every ballot cast for its player awards. Each of its 211 member associations sends up to three jurors, the head coach and the captain of its national team plus one journalist, and each juror ranks three players, who receive five, three and one points. We assembled all 15,385 ballots cast for the men's and women's awards from 2010 to 2025, spanning 32 contests in which the same three professions judge the same nominees. France Football's Ballon d'Or, awarded in the same years to largely the same players, works differently: one journalist per country ranks five players (ten from 2024). Its 706 ballots from 8 contests, cast by smaller and more European juries of 47 to 176, give us a second jury design for free.

The published ballots survive auditing. FIFA's published result breakdowns give each juror group's points for the leading nominees in 2023 and 2024, and all 36 recomputed values agree exactly \cite{fifa2023rules}. For the earlier FIFA years, when no fan vote existed, every nominee's published share of the vote is reproduced by our parsing to within a quarter of a percentage point. For the Ballon d'Or, recomputed totals equal the published totals with three documented single-point exceptions. Two historical quirks are handled explicitly: captains who are themselves nominees may not vote for themselves and are removed from their own ballots, and in 2010 votes by captains and coaches for players of their own national team were void, so those pairings are excluded from that year's menus.

Two features of the data turn out to matter later. Since 2016 fans have cast a quarter of FIFA's vote and their ballots are not published, so for those contests we study the standing of the three jury groups under FIFA's scoring rule of the year; the jury's leader went on to win the award in 17 of those 20 contests. And many national teams are coached by foreign nationals: the coach's own nationality and the country the coach works for differ on 1,246 of the 2,727 ballots whose coach we could identify. Those coaches offer a rare natural experiment on what favoritism is really about.

\section{Measuring favoritism on a ranked ballot}

A ballot is a ranking, and rankings have a convenient statistical structure. Think of each nominee in a contest as a horse in a race: every juror watches the horses run, and the order in which they cross the juror's personal finish line is the ballot. Formally, juror $i$ of profession $r(i)$ assigns nominee $j$ in contest $c$ a utility
\begin{equation*}
u_{icj} \;=\; \alpha_{cj} \;+\; \beta_{\mathrm{nat},r(i)}\, x^{\mathrm{nat}}_{icj} \;+\; \beta_{\mathrm{conf},r(i)}\, x^{\mathrm{conf}}_{icj} \;+\; \beta_{\mathrm{lea},r(i)}\, x^{\mathrm{lea}}_{icj} \;+\; \varepsilon_{icj},
\end{equation*}
and ranks the nominees in decreasing order of utility. This is the Plackett Luce model, a workhorse of rank analysis \cite{luce1959,plackett1975,beggs1981,allison1994}. Its statistical machinery is well developed: efficient likelihood based and Bayesian computation go back two decades \cite{hunter2004,guiver2009,caron2012}, Bayesian versions accommodate covariates and heterogeneous rankers \cite{liyiliu2022}, and mixtures of the same model have been used to uncover voting blocs within electorates \cite{gormley2008,mollica2017}, an idea our bloc level findings connect to directly. The term $\alpha_{cj}$ is the nominee's appeal, everything that makes the nominee attractive to every juror in that contest: talent, form, fame. The indicators $x$ record whether juror and nominee share a nationality, a confederation, or the country of one of the nominee's clubs, and each coefficient $\beta$ is a similarity bonus: how much extra preference a nominee gets from jurors who share that tie, over and above what other jurors give the same nominee in the same vote. Because appeal is estimated separately for every nominee in every contest, a bonus cannot be driven by the nominee simply being good. Standard errors are clustered by association, allowing the ballots of one association to correlate freely across professions and years.

The bonuses pass a battery of sanity checks. Shuffle juror countries among the jurors of each contest and profession, which breaks every real tie between jurors and nominees, and the compatriot bonus vanishes (mean $-0.01$ over 100 shuffles); shuffle within confederations and the compatriot bonus vanishes while the regional one survives, exactly as it should if the two operate at different levels. Refit the model under 42 combinations of fixed effects, samples and outcomes, and all 126 compatriot coefficients and 82 of 84 regional ones keep confidence intervals above zero. Jurors also do not mark down the rivals of a compatriot contender, so the bonuses read as favoritism rather than sabotage. And in the 2011 to 2014 sample studied by Coup\'e and coauthors, our journalists' compatriot odds ratio of 9.9 nearly reproduces theirs, while our larger estimates for coaches and captains reflect the richer appeal structure \cite{coupe2016wp}.

\begin{figure}[t]
\centering
\includegraphics[width=\textwidth]{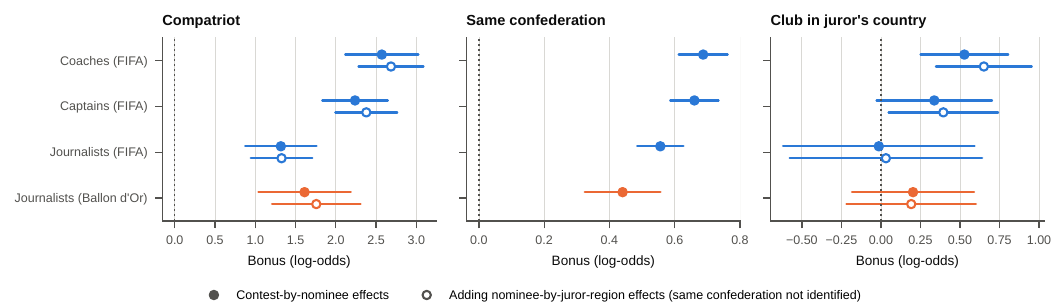}
\caption{Similarity bonuses by profession, in log odds per ranking step, with 95\% confidence intervals. Filled points come from the main model; open points add a separate appeal for each nominee among jurors of each region, which absorbs any regional taste for a nominee's style of play. FIFA's coaches, captains and journalists in blue, Ballon d'Or journalists in orange.}
\end{figure}

Every profession favors compatriots (Figure~2). A compatriot's odds of being chosen at any ranking step are multiplied by 13.1 for coaches, 9.4 for captains, 3.7 for FIFA's journalists and 5.0 for the Ballon d'Or's journalists, with larger bonuses in the men's award than in the women's. Regional affinity is real but gentler: jurors favor nominees from their own confederation by factors of only 1.6 to 2.0, and coaches add a factor of 1.7 for players whose clubs are in their country. Read as individual psychology, the message seems clear. Compatriot favoritism is the strongest effect in the room, so it must be the one that matters. The rest of this article argues that this reading is exactly wrong.

\section{Replaying the votes with the bias switched off}

To ask whether the bonuses changed any actual result, we replay every contest. Under the horse race view, a ballot is the order in which random clocks ring, and the observed ranking pins down those clock times up to a draw from a known distribution. We can therefore redraw each juror's randomness consistent with what the juror actually did, subtract the estimated bonuses, let the same clocks ring again, and recount the votes. Observed and replayed ballots share all their randomness and differ only through the bonuses, so a changed winner is attributable to favoritism alone. Each replay probability also averages over the uncertainty of the estimates; tested on simulated data, including noise the model does not assume, the estimated number of changed leaders came within about a tenth of the truth.

\begin{figure}[t]
\centering
\includegraphics[width=0.6\textwidth]{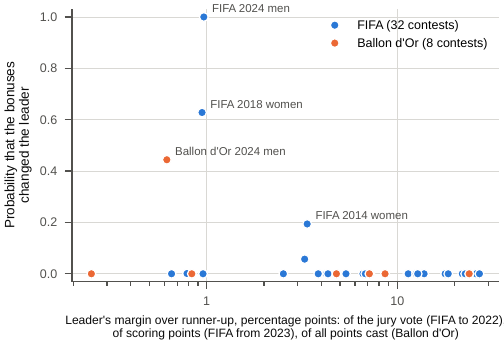}
\caption{When the bonuses changed the leader. For each of the 40 contests, the probability that the winner would have differed had every juror ranked the nominees without the similarity bonuses, against the winner's observed margin. Only close contests are exposed, and all of them are recent.}
\end{figure}

Across all 40 contests, removing the bonuses changed the jury leader with probability above one half in two contests and above one tenth in four, an expected 2.3 changed leaders in total (Figure~3). Here is the surprise. Every one of those changes came from the regional bonus. Removing only the compatriot bonuses changed no leader at all, and neither did removing only the league bonuses, whereas removing only the regional bonus reproduced the full 2.3. The strongest tie in the data, exercised by at most three compatriots per nominee on FIFA's juries and one on France Football's, never once tipped a contest. The weaker tie, exercised by a quarter of all juror nominee pairs, did all the damage. The regional bonus also differs between regions, from 0.45 in log odds for CONCACAF jurors and 0.53 for UEFA jurors up to 1.18 for African jurors, and with region specific bonuses the expected number of changes is 1.7, still all from the regional term. One might worry that the regional bonus merely measures a regional taste for a style of play rather than a tie between juror and nominee. A model that lets every nominee appeal separately to jurors of each region absorbs all such taste, and the compatriot and league bonuses that survive then change no leader, so the outcome level findings are not an artifact of regional style.

The contests where the bonuses mattered were close, and they make good stories. In FIFA's 2024 men's award, Rodri won the coaches' vote by 23 points and the journalists' by 5, while Vin\'icius J\'unior won the captains' by 244. Under FIFA's current rule, which converts ranks within each group to a fixed scale of scoring points, Rodri led the combined jury standing by 0.97 percentage points. Without the bonuses, Vin\'icius J\'unior would have led instead, with probability 1.00. The twist is that under the rule FIFA used until 2022, which scores points rather than ranks, Vin\'icius J\'unior would have led by 4.15 points and the bonuses would have changed nothing. Same ballots, same voters, same bias, different arithmetic, different verdict. In the 2018 women's award, Ada Hegerberg led the jury standing by 0.95 percentage points, and without the bonuses another nominee, most often Megan Rapinoe, would have led with probability 0.63; that race was close enough that the ballots of any single association could matter, the probability ranging from 0.47 to 0.98 as associations are left out one at a time. In the 2014 women's award the probability was 0.19. In the 2024 Ballon d'Or, decided by 0.62 percentage points, removing the bonuses shrank the expected margin to 0.03 and flipped the winner with probability 0.44, though a model with region specific bonuses, which fit the European jurors a smaller regional bonus, puts it at 0.10. Favoritism does not always help the favorite, either: in the 2019 Ballon d'Or the bonuses worked against the winner, whose margin would have grown from 0.25 to 1.53 points without them. Both verdicts agree with our deeper Bayesian reanalysis of these two contests \cite{huwang2026bdo}. The bonuses tilt close races in whichever direction the blocs happen to lean.

How much of this depends on the model? It is worth being precise, because our replay probabilities are model based counterfactuals, not observed interventions. Three layers of evidence bear on the question. First, the procedure itself is validated: on synthetic contests generated from the fitted model, including contests generated with a noise distribution the model does not assume, the estimated number of changed leaders came within about a tenth of the truth, so the replay machinery recovers known answers. Second, the individual contest probabilities are genuinely specification sensitive. The 2024 Ballon d'Or flip probability is 0.44 under a common regional bonus and 0.10 when the bonus may differ by region; the 2018 women's probability ranges from 0.47 to 0.98 as single associations are left out; and a model that absorbs all regional taste into nominee appeal leaves no contest with a flip probability above 0.03. We therefore do not read any single contest's number as decisive. Third, the qualitative conclusions are stable in ways the individual numbers are not. Across common and region specific bonuses, two noise distributions, 42 combinations of fixed effects, samples and outcomes, and the leave one out refits, three statements survive: the compatriot bonus alone never changes a leader, the regional bonus accounts for every change that does occur, and the changes concentrate in the closest contests. Our headline claims are stated at that level.

\section{Designing better juries}

If you ran an award, what should you do about this? We answered the question by simulation: build juries that vote with the estimated appeals and bonuses, run each vote twice with the same random draws, once with the bonuses and once without, and record how often the winners differ. Exposure, it turns out, is governed by four ingredients: how strong the preference is, how many jurors share it, how unevenly it is distributed between the two leading candidates, and how close the race is.

\begin{figure}[t]
\centering
\includegraphics[width=\textwidth]{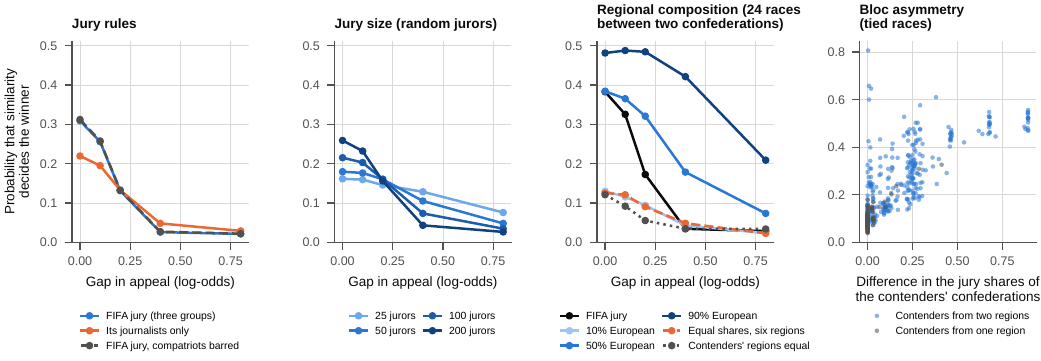}
\caption{Jury design. Probability that similarity bonuses decide the winner of a simulated vote, as the gap in appeal between the two leading nominees grows. (a) FIFA's actual jury, its journalists alone, and FIFA's jury with compatriot votes barred. (b) Random juries of different sizes. (c) Races between two confederations under different regional compositions of the jury. (d) The probability at a tie against the difference between the two contenders' bloc shares on the jury.}
\end{figure}

Closeness dominates everything (Figure~4). Averaged over FIFA's contests, the bonuses decided 31\% of tied races but only 3\% of races with a modest appeal gap between the top two nominees. After closeness comes imbalance. A regional bonus pulls two nominees apart only to the extent that their regions are unequally represented on the jury: the bonuses decided 39\% of tied races when the difference between the contenders' bloc shares exceeded 25 percentage points, 25\% when it lay between 5 and 25 points, 17\% when it fell below 5 points, and 10\% when both nominees came from the same confederation, with a correlation of 0.70 between the imbalance and the probability across all simulated juries. In the 24 contests whose two leading nominees came from different confederations, FIFA's jury decided a tied race by the bonuses with probability 0.38. Give each confederation equal numbers of jurors and the probability falls to 0.13; simply shrink the larger contender's bloc to the size of the smaller one and it falls to 0.12. The European share of the jury acts through the same channel: in juries of 100 with 10\% Europeans the contenders' blocs differed by 5 points on average and the probability was 0.13, while with 90\% Europeans the blocs differed by 64 points and the probability rose to 0.48. The same ordering survives region specific bonuses (0.35 for FIFA's jury, 0.12 with balanced contenders) and a different noise distribution (0.37 and 0.11).

Jury size, the variable organizers most often fiddle with, matters least: random juries of 25 and 200 jurors gave 0.16 and 0.26 for a tied race, because chance makes races close in small juries but rarely masks a bloc wide shift in large ones. And the standard safeguard fails completely: barring jurors from ranking their compatriots left the probability unchanged at 0.31, moving no contest by more than 0.07. Composition, not size, is the lever. Replaying the close 2019 and 2024 Ballon d'Or races under juries of FIFA's design rather than France Football's own raised the exposure from 0.18 and 0.23 to 0.32 and 0.29, consistent with FIFA's jurors holding the larger regional bonuses. Even the same shortlist, voted on the same year, is safer under a jury whose blocs are balanced.

\section{Coaches abroad: favoritism follows the job, affinity does not}

What is this favoritism made of? The 1,246 ballots cast by coaches working for a country that is not their own separate a juror's passport from a juror's paycheck. The answer is striking (Figure~5). Foreign coaches favored players of the country they coached by 2.99 in log odds (95\% interval 2.20 to 3.79), as much as domestic coaches favored their compatriots, 2.36 (1.84 to 2.87). The coach's own nationality also mattered, but distinctly less, at 1.47 (1.24 to 1.70), and the gap between the coached country and the coach's own nationality appears in both the men's and the women's award. In the women's award, which supplies most of the relevant ballots, foreign coaches favored the coached country more strongly than domestic coaches favored theirs (2.82 against 1.13). National favoritism is not written in the passport; it follows the job. Part of this need not be bias at all: coaches plausibly know their own players best, and familiarity carries decision relevant information \cite{li2017}.

\begin{figure}[t]
\centering
\includegraphics[width=0.7\textwidth]{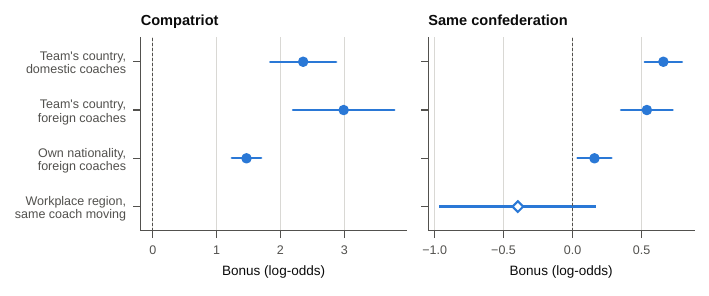}
\caption{Foreign national-team coaches. Bonuses for nominees of the country a coach works for (domestic and foreign coaches) and of a foreign coach's own nationality, for nationality and confederation. The last row is the bonus for the confederation of work estimated within coaches who voted from two or more confederations. Bars are 95\% confidence intervals clustered by juror association.}
\end{figure}

Regional affinity tells a different story. Foreign coaches of men's teams did favor nominees from the confederation in which they worked more than nominees from their confederation of origin, but that comparison is between coaches, and coaches who take jobs in a region may already favor it. The cleaner test is within coach: among the 16 coaches who cast ballots from two or more confederations, whose lasting regional tastes are absorbed by coach specific fixed effects, the bonus for the current region of work was $-0.40$ (standard error 0.29). We find no evidence that regional affinity follows a coach from one posting to the next, although the interval cannot rule out a workplace bonus of up to about a quarter of the between coach value. Players who changed clubs between leagues offer a similarly inconclusive test, because few nominee and juror country pairs change state. National loyalty is portable and professional; regional affinity is neither. The two preferences look similar on a ballot but seem to be manufactured by different processes.

\section{What juries everywhere should learn}

Expert juries are usually designed around the strongest individual tie between an evaluator and a candidate. Sixteen years of football ballots suggest that this focus misses the preferences most capable of changing collective outcomes. A strong preference exercised by few jurors changed nothing; a weaker preference shared by a bloc tipped several close races. Individual conflicts and collective exposure are distinct design problems, and solving the first does not solve the second.

The practical implication is conditional but concrete. Within our simulations, balancing the leading contenders' regional blocs cut the probability that favoritism decides a tied race from 0.38 to 0.12, while barring compatriot votes did almost nothing. Balanced representation does not eliminate anyone's preferences; it lets them cancel, and only imperfectly when bonuses differ across regions. Organizers must weigh that benefit against expertise, informational diversity and plain feasibility. Our estimates also describe departures from what other jurors think, not from some objectively correct ranking, and regional affinity may be taste or information rather than favoritism. The replays are model based counterfactuals rather than real interventions, and as the previous section details, probabilities for individual contests are sensitive to specification, even though the ordering of reach over strength is not. Coach nationality could be established for 2,727 of 5,364 coach ballots with coverage varying across regions, fan votes meant that two of the clearest cases did not change the actual award, and football's unusually transparent awards are not a representative sample of expert panels. Whether the same mechanism operates in grant review, hiring or legal decision making is one of the more interesting open questions in the statistics of evaluation, and the toolkit used here, a ranking model, ballot conditioned counterfactuals and design simulations, travels to those settings unchanged.

One lesson needs no qualification. Publishing individual ballots, as FIFA and France Football do, is what makes effects like these visible at all; transparency turned a trophy vote into a natural experiment on jury design. All data and code behind every number in this article, from parsing the ballot files to the models, replays and simulations, are publicly available \cite{huwang2026code}.

\end{document}